\documentclass[sigconf,screen]{acmart}
\usepackage{multirow}
\usepackage{multicol}

\usepackage{fancyhdr,graphicx,amsmath,amssymb}
\usepackage[ruled,vlined]{algorithm2e}
\AtBeginDocument{%
  \providecommand\BibTeX{{%
    \normalfont B\kern-0.5em{\scshape i\kern-0.25em b}\kern-0.8em\TeX}}}
\setcopyright{acmcopyright}
\copyrightyear{2023}
\acmYear{2023}
\acmDOI{XXXXXXX.XXXXXXX}

\acmConference[ACM BlockSys]{Fifth ACM International Workshop on Blockchain-enabled Networked Sensor Systems}{November 12, 2023}{Istanbul, Turkiye}
\acmConference[BlockSys]{Fifth ACM International Workshop on Blockchain-enabled Networked Sensor Systems, November 12, 2023, Istanbul, Turkiye} 
\acmPrice{15.00}
\acmISBN{XXX-X-XXXX-XXXX-X/XX/XX}

\usepackage{etoolbox}
\makeatletter
\patchcmd{\maketitle}{\@copyrightspace}{}{}{}
\makeatother
\begin{document}

\title{ANKA: A Decentralized Blockchain-based Energy Marketplace for Battery-powered Devices}

\author{Burak Can Sahin, Abdulrezzak Zekiye, Oznur Ozkasap} 

\email{{ buraksahin20, azakieh22, oozkasap }@ku.edu.tr}
\affiliation{%
  \institution{Department of Computer Engineering, Koç University}
  \city{Istanbul}
  \country{Türkiye}
  \postcode{34450}
}


\begin{abstract}

For the purpose of enabling, democratizing, and reducing the fees of peer-to-peer energy trading for battery-powered devices, we propose ANKA as a fully decentralized energy marketplace for peers with battery-powered devices. ANKA utilizes state-of-the-art technologies, namely blockchain, smart contracts, and decentralized applications. Within this marketplace, users who possess surplus energy actively offer their excess energy for trading. Concurrently, consumers can readily explore the energy options available and make purchases according to their individual preferences while taking into consideration the location of the offered energy and voltage compatibility. In addition, we provide a comparison between a centralized traditional market and our proposed solution, identifying that the cost of deploying and operating ANKA is less than the centralized approach. We also position ANKA in comparison to the recent blockchain-based decentralized energy marketplaces by considering the metrics of blockchain type, scope, trading entities and the presence of third parties. 
%

\end{abstract}





\maketitle

\section{Introduction}

\begin{figure}[h]
  \centering
  \includegraphics[width=0.9\linewidth]{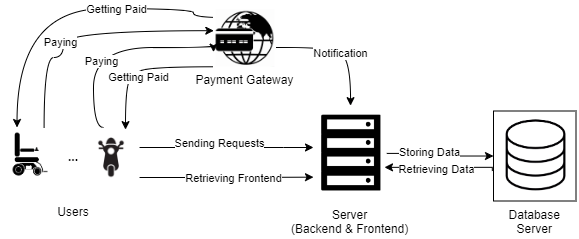}
  \caption{Centralized marketplace architecture}
  \label{fig:centralized_market}
\end{figure}

An energy marketplace aims to connect energy sellers and buyers so that energy sellers can transfer their excess energy to the buyers and get paid for it. An example of a centralized marketplace for electricity trading in smart grid neighborhoods can be seen in \cite{ilic2012energy}. In Fig. \ref{fig:centralized_market}, a traditional centralized marketplace is depicted that consists of a backend software, a database, a frontend part which is a user interface that lets the user interact with the system, and a payment gateway connected to a centralized third-party service. Usually, the backend and frontend are hosted on a server, and the database could be on the same server or a separate one. Such a marketplace has multiple points of failure, such as the backend server, the database, and the payment gateway. In addition, those centralized marketplaces are not democratized since they are controlled by third parties and that might lead to increased fees either for the servers to host the backend or the database. Other fees might exist for listing an item for selling in such markets or in the form of fees taken as a percentage from each selling operation. 

Blockchain, decentralized applications and smart contracts provide solutions to the aforementioned problems. Blockchain is a distributed and decentralized ledger that is persistent, anonymous, and auditable \cite{zheng2018blockchain}. Blockchains can be classified into three main types according to the ability to join them. (1) Public blockchains are open and decentralized, (2) private blockchains are controlled by a central authority, and (3) consortium blockchains are private ones with multiple authorities controlling them instead of only one. Smart contracts are blockchain-hosted programs that shift the trust from third parties to the code and are capable of saving costs \cite{zheng2020overview}. Decentralized application (dApp) represents the user interface that we can use to interact with the smart contract where it serves as the frontend part of a centralized market, except for having the data and computation provided by the blockchain \cite{metcalfe2020ethereum}. The blockchain and smart contracts can replace the backend server, the database, and the payment provider, while the decentralized application replaces the traditional frontend, as a result, we get a decentralized marketplace.

To recharge the battery of a powered vehicle, users have to search for charging stations. However, there may not be any charging stations nearby which could make the device unusable and affect the user's daily life. In a scenario where a powered wheelchair user could not find any docks to charge their device, it would reduce the user's mobility. As a solution, we propose ANKA, a decentralized peer-to-peer (P2P) energy marketplace where users can search for near energy offers, placed by other peers, based on their needs and match. ANKA reduces the dependency on charging docs since even if there are no charging stations, it would be possible to find other battery-powered device users (such as scooters and wheelchairs) that would be available for trading.

The contributions of this study are as follows:
\begin{enumerate}
    \item The proposal and implementation of a fully decentralized marketplace, namely ANKA, for peer-to-peer energy trading between battery-powered devices. Unlike the prior work Joulin \cite{perk2020joulin}, our proposed marketplace is fully decentralized without any single point of failure due to the usage of only blockchain, smart contracts, and decentralized applications. Compared to the works in \cite{tkachuk2023towards, mezquita2022towards} that used a permissioned blockchain, ANKA utilizes a public blockchain which means our system is not controlled by any entity and provides full decentralization.
    \item ANKA has no third parties thus there are no extra fees or taxes to be paid, unlike the not fully decentralized solution \cite{perk2020joulin}. Researchers in \cite{tkachuk2023towards} had a third party to govern the market; however since our system is not connected to any grid and power exchange happens really peer-to-peer, ANKA has no such entity. 
    \item Our proposed marketplace ANKA considers a free-pricing policy where sellers put the price they want and buyers decide themselves whether to buy or not. In \cite{khorasany2021lightweight, perk2020joulin}, the proposed systems tend to force a pricing policy or do an automatic matching between peers.
    \item The proposed ANKA system takes into consideration any battery-based device (such as powered wheelchairs, electric scooters, electric vehicles, and power banks) to trade the energy from it where the seller declares the battery voltage when making a listing and the buyer will have the choice to buy energy from voltage-compatible devices to his own. The proposed systems in \cite{khorasany2021lightweight, perk2020joulin, tkachuk2023towards} are either limited to utility power or consider only one type of devices, mostly electric vehicles.
\end{enumerate}

\section{Literature Review}
In this section, we provide an overview of the latest blockchain-based decentralized energy marketplaces and compare them to our solution, ANKA.
In Table \ref{tab:summary}, we present an overview the recent blockchain-based decentralized energy marketplaces in comparison to ANKA. The table outlines the utilized type of blockchain, the extent of the proposed solution (scope), the involved trading entities, and the presence of third parties or potential single points of failure.

In \cite{perk2020joulin}, a blockchain-based peer-to-peer energy trading system named Joulin was proposed. Some of the components of Joulin are public blockchain, a matching service, an event watcher, and a database. Joulin is not fully decentralized since a server was used to do the matching between sellers and buyers and a centralized database exists to store a copy of the smart contract's data. As a result, there is the server as a single point of failure along with the database, and also an increased cost since there are the server's fees and the blockchain ones. More than that, the entity that deployed the system is considered as an owner and can apply taxes. 
  
A blockchain-based framework for peer-to-peer energy trading in a smart grid has been proposed in \cite{khorasany2021lightweight} where the distances between peers are considered in pricing and a reputation factor has also been introduced for each agent. Producers and consumers advertise their needs through the grid operator, which stores the requests in a centralized database.
In \cite{tkachuk2023towards}, a peer-to-peer blockchain-based energy marketplace was introduced. Their system addressed privacy, trust, and governance issues by using a permissioned blockchain and a regulatory third party. This regulator ensures certified usage of renewable energy sources (RESs) and oversees energy transactions for accurate electricity mapping.  
 

The collection of loads, distributed generation resources (DES), and energy storage systems (ESS) in a specific place defines what is called a microgrid \cite{garg2018overview}. A platform for the creation of an automated peer-to-peer in-microgrid energy market was proposed in \cite{mezquita2022towards}. To increase privacy and reduce cost, a permissioned blockchain was used. The platform allows for choosing the price-negotiation algorithm while preserving anonymity. 

The usage of fungible tokens (FT) and non-fungible tokens (NFT) with Hyperledger Fabric for energy trading has been proposed by researchers in \cite{karandikar2021blockchain}. NFTs were used to represent assets with value and unique information or identifier while FTs were used to represent assets with value only. The system involves different participants and tokens with specific roles. Prosumers can buy and sell energy from other peers and thus earn an NFT as a reward. Electric vehicle (EV) owners can buy energy to charge their vehicles. They can also rent out their vehicle's battery and get an NFT reward as well. The entities in the proposed system are prosumers, electric vehicle owners, power companies, and storage providers. Prosumers and EV owners can join the system to sell or buy energy, earning reward tokens through tasks like demand response. They can also access the Storage Provider via the platform to store or withdraw excess energy. Power Companies can engage customers by offering reward tokens for demand response tasks. Gamification tokens are used to incentivize tasks like conserving energy on hot days.

\begin{table*}
  \caption{Summary of decentralized blockchain-based energy marketplace studies}
  \label{tab:summary}
  \begin{tabular}{l|l|l|l|ll}
    \toprule
    Ref & Blockchain Type & Scope & Entities & Third Parties/Single Point of Failure \\
    \midrule
    \cite{perk2020joulin} & Public & Smart Grid & \parbox[l]{3.5cm}{Producers, Consumers} &  Server, database \\

    \hline
    \cite{khorasany2021lightweight} & Public & Smart Grid & \parbox[l]{5cm}{Producers, Consumers, Grid} & Advertisement Database, Grid Operator \\
    \hline
    \cite{tkachuk2023towards} & Permissioned & - & \parbox[l]{6.3cm}{Producers, Consumers, Providers, Regulator} & Regulator, Network Owners \\
    \hline
    \cite{mezquita2022towards} & Permissioned & Microgrid &  \parbox[l]{5cm}{Producers, Consumers, Main grid} & Network Owners   \\
    \hline
    \cite{karandikar2021blockchain} & Permissioned & Microgrid & \parbox[l]{6.3cm}{Producers, Consumers, EV Owners,\\ Power Companies, Storage Providers}  & Network Owners\\
    \hline
    ANKA & Public & Global & \parbox[l]{3.7cm}{Battery-powered devices} & None  \\
  \bottomrule
\end{tabular}
\end{table*}

\section{System Description}

According to the classifications of blockchain-based peer-to-peer energy trading systems in \cite{ali2020cyberphysical}, ANKA can be considered as infrastructure-based energy trading platform. The peers in ANKA are sellers and buyers. Sellers are owners of battery-based devices with an excess of power and they are willing to trade it with others. Buyers are also owners of battery-based devices with the need for power. A decentralized energy marketplace can connect the sellers and buyers to trade energy between themselves. Like any other marketplace, sellers list their product details and buyers can browse the listed products and select the suitable offer to buy. Products in our marketplace are energy offers with information such as the amount of offered energy, the voltage type (e.g. 9V, 12V, 24V), the price, and the location. Buyers can filter out energy offers based on the requested voltage type and then select an offer according to their own criteria, such as location, price, and offer energy amount. Once a suitable energy offer is found by the buyer, the buyer can navigate to the place of the offered energy, charge their battery-based device from the device that offered the energy, and complete the transaction over the blockchain.

\begin{figure}[h]
  \centering
  \includegraphics[width=\linewidth]{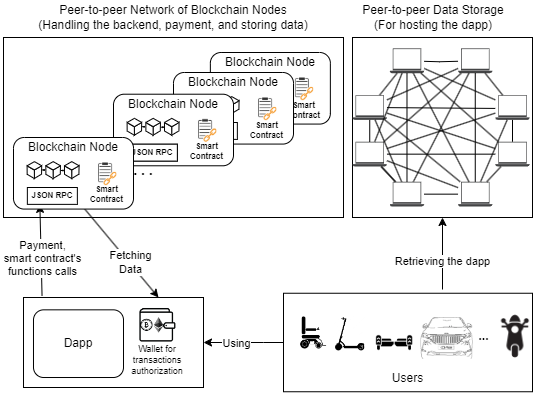}
  \caption{Overview of the proposed ANKA decentralized energy marketplace for battery-powered devices}
  \label{fig:decentralized_market}
\end{figure}

An overview of the ANKA is depicted in Fig. \ref{fig:decentralized_market}. The blockchain itself plays the role of a server in a centralized marketplace. The smart contract, on the other hand, is the equivalent of the backend code, the database, and the payment gateway in the traditional marketplace. The decentralized application (dApp) is the user interface to use the system and it gets and writes data to the smart contract. The dApp can be hosted on a server. However, that contradicts the purpose of our system, which is making it fully decentralized with no third parties at all. For that reason, we proposed the usage of decentralized storage systems such as InterPlanetary File System (IPFS) \cite{benet2014ipfs}.

Each user in the blockchain is identified by his unique public address. The user has also his own private key that corresponds to his public address. The private key is used to authenticate any transaction on the blockchain (such as a call to a smart contract function or transferring tokens). Cryptocurrency wallets are a user-friendly way to manage user's public and private keys and allow the user to do transactions in an easy way \cite{suratkar2020cryptocurrency}. 

In our ANKA system, a wallet is needed to make users able to use the decentralized application. After connecting to the blockchain via the wallet, the decentralized application connects to the smart contract and prompts the user to register in the system with a name that will be used as a friendly way to display the owner of an energy offer. After the registration process, the user will be able to add an energy offer to the system by specifying the necessary information about the offer such as the price, energy amount, and voltage type. It is worth mentioning that the cryptocurrency wallet works by connecting itself to the blockchain using a JSON-RPC API, which means JavaScript Object Notation-Remote Procedure Call Application Interface. In the same way, the dApp uses web3 libraries that connect to the blockchain using JSON-RPC API.

Selecting an energy offer to buy is an important step and should be considered in a way that minimizes the gas cost. Logically, a buyer will need to check the energy offers that are near his location. Using the buyer's latitude and longitude to get the available energy offers within a specific diameter can be done by filtering the offers through iteration as we can see in Algorithm \ref{alg:diameter_filtering}. This algorithm can be implemented either in the smart contract or on the client side. However, both solutions have problems. Executing the process on the smart contract will lead to increased gas fees. On the other hand, filtering the offers on the client side might take time if the offers' number is big and will lead to slowness in the decentralized application. To solve this problem, we introduced a postal code location search method where the energy posts are stored in dictionaries with postal codes as keys and energy offers as values. As a result, users will be able to access the energy posts in their area easily without any increased gas fees or increased processing on their end. Considering energy offer dates; checking various days is unfeasible due to battery discharge rates. The system enables local trading of excess energy using streamlined date-based dictionaries, reducing complexity, cost, and client burden. Thus, the smart contract retrieves energy listings by accessing date-based and postal code dictionaries. The user views and selects offers based on voltage, energy amount, and location.

\begin{algorithm}[htp]
\SetAlgoLined
\SetKwInOut{Input}{Input}\SetKwInOut{Output}{Output}
\KwResult{filteredEnergyOffers}
\Input{buyer's location, diameter, energyOffers}
\SetKwInOut{r}{tt}
    $filteredEnergyOffers \leftarrow []$

    \ForEach{$energyOffer$ $\in$ $energyOffers$}{
        \If{distance between energyOffer and buyer's location < $diameter$}{
            add $energyOffer$ to $filteredEnergyOffers$
        }
    }    
  \caption{Fetching energy offers within a specific area using the buyer's latitude and longitude}
  \label{alg:diameter_filtering}
\end{algorithm}


\section{Discussion}

In centralized markets, usually, the owner pays an annual fee for the hosting service. On the other hand, the smart contract will be deployed once, in other words, it is a one-time fee. Making a product listing in a centralized market is subject to the market owner. Using a smart contract with no fees paid for any third party, there will be only gas fees that depend on the listing code. The gas fees' fiat currency equivalent changes depending on the blockchain network usage. Finally, when making a sale in traditional centralized markets, the owner gets a percentage out of the paid price and the payment gateway gets a fee as well. In ANKA, there is no percentage deducted from the paid price, thus 100\% of the money gets transferred directly to the other party and some negligible fees get paid to the network. In addition, gas price exists and will be paid by the buyer, unlike centralized solutions where the buyer usually pays no extra fees. 

Table \ref{tab:cost_comparision} shows a comparison between the cost of centralized marketplaces and ANKA. Server cost was taken as the average of the lowest hosting plan in four different hosting companies, namely Godaddy, Hostinger, IONOS, and Bluehost. Regarding payment gateway fees, we considered the average of the deducted percentage by PayPal, Braintree, Stripe, and Razorpay and neglected the additional fixed cost. The fees of product listing, selling fees, and buying fees were calculated as the average of corresponding fees from Hepsiburada, Amazon Türkiye, n11, and Trendyol. The fees regarding ANKA were taken during testing our system and recorded in USD. The price of 1 Gwei (giga wei) at the time of writing this paper was approximately 0.00000164534 USD \cite{coinmarketcap}.

\begin{table}
  \caption{Cost comparison between a centralized marketplace and our proposed decentralized solution}
  \label{tab:cost_comparision}
  \begin{tabular}{l|l|l}
    \toprule
    \parbox[c]{1cm}{Market Type}&Technology&\parbox[c]{1cm}{Approximate\\Cost}   \\
    \midrule
    \multirow{5}{*}{\parbox[c]{1.6cm}{Centralized Market}} & \parbox[c]{4cm}{Server (frontend, backend, database)} &\$52.58\ yearly\\
    \cline{2-3}
     & Payment Gateway  &   2.57\% \\
     \cline{2-3}
     & Product Listing &   0\\
     \cline{2-3}
     & Selling Fees &  15.58\%\\
     \cline{2-3}
     & Buying Fees & 0\\
     \hline

     \multirow{4}{*}{\parbox[c]{1.6cm}{Decentralized Market}} & \parbox[c]{4cm}{Deployment (smart contract)} &  \parbox[c]{1.8cm}{ \$5.40 (once)}\\
    \cline{2-3}
     & Payment Gateway & N.A.\\
     \cline{2-3}
     & Product Listing & \$0.88 \\
     \cline{2-3}
     & Selling Fees & 0\%\\
     \cline{2-3}
     & Buying Fees & \$0.12 \\ 
  \bottomrule
\end{tabular}
\end{table}

Though the ANKA has no single point of failure, and it seems to have less fee, it is less scalable than the centralized one due to consensus protocol. Blockchains such as Lightchain \cite{hassanzadeh2019lightchain} and Solana \cite{yakovenko2018solana} are faster and more scalable than Ethereum; however, a thorough investigation and comparison with a centralized solution should be made. In case decentralization is not needed or there will be a third party that has control over the market, it is better to consider a centralized marketplace approach.

\section{Conclusion and Future Work}

We propose ANKA, a fully decentralized energy marketplace for battery-powered devices, that utilizes state-of-the-art technologies of blockchain, smart contracts, and decentralized applications. 
ANKA facilitates the exchange of power between sellers, who possess surplus energy in their battery-based devices, and buyers seeking power for their devices. The marketplace employs a blockchain-based system with smart contracts for backend functionality and a user-friendly decentralized application for interaction, hosted on a decentralized peer-to-peer storage. An approach involving postal code-based energy post storage optimizes user access to nearby energy offers without gas fee escalation or application slowdowns. Additionally, a hierarchical date-keyed structure further refines energy and offers retrieval. Through these innovations, our system provides an efficient and effective means for buyers and sellers to engage in energy trading while addressing location, date, and device-specific criteria.

For future work, simulations of the proposed system to measure its scalability and efficiency along with a comparison between other possible networks could be done. In addition, privacy issues and anti-malicious behavior mechanisms could be considered. One form of malicious behavior is making a fake listing. Though listing is not free, its cost is not high enough to prevent such behavior. 

In centralized markets, such behavior is prevented by validating the sellers' identities by the market owner, and in case a malicious behavior is detected, the buyer can be compensated.  Additionally, the malicious seller might receive negative feedback and potentially be banned from the marketplace for repeated misconduct. However, these measures are impractical in the context of ANKA because (a) ANKA operates without any involvement of third parties, which makes it impossible to verify the identities of its participants, (b) as transactions are conducted on a peer-to-peer basis, the possibility of compensation is eliminated. Nevertheless, within the ANKA system, the process involves buyers acquiring energy directly from sellers. Consequently, if any malicious activity is identified, the buyer will withhold payment. (c) Blockchain users remain anonymous, rendering attempts to implement review or ban systems ineffective. This is because a banned user or an individual with negative feedback can effortlessly rejoin ANKA using a fresh public address.


\begin{acks}
This work was supported in part by TUBITAK 2247-A Award 121C338.
\end{acks}

\bibliographystyle{ACM-Reference-Format}
\bibliography{sample-base}

\end{document}